\documentclass[conference]{IEEEtran}
\usepackage{amsmath, amssymb,multicol, graphicx,gensymb, flushend}

\IEEEoverridecommandlockouts

\begin{document}

\title{Activation of Polylactic Acid and Polycarbonate Surfaces with Non-Thermal Plasma}
\author{\IEEEauthorblockN{Jairo Rondón\IEEEauthorrefmark{1}, 
Ginger Urrutia\IEEEauthorrefmark{2}, Angel Gonzalez-Lizardo\IEEEauthorrefmark{3}}
\IEEEauthorblockA{\IEEEauthorrefmark{1}Biomedical \& Chemical Engineering Departments,\\
Polytechnic University of Puerto Rico, San Juan, Puerto Rico, USA\\jrondon@pupr.edu}
\IEEEauthorblockA{\IEEEauthorrefmark{2}Biomedical Engineering Department,\\
Polytechnic University of Puerto Rico, San Juan, Puerto Rico, USA\\urrutia\_127357@students.pupr.edu
\IEEEauthorblockA{\IEEEauthorrefmark{3}Department of Electrical and Computer Engineering and Computer Science,\\
Polytechnic University of Puerto Rico, San Juan, Puerto Rico, USA\\agonzalez@pupr.edu}
}}

\maketitle

\begin{abstract}

Non-thermal plasma (NTP) surface activation has become a powerful and versatile strategy to engineer the interfacial properties of biomedical polymers whose intrinsic hydrophobicity limits their biological performance. In polymers such as polylactic acid (PLA) and polycarbonate (PC), NTP promotes the controlled incorporation of polar functional groups, increases surface energy, modifies dielectric behavior, and generates micro-roughness that collectively enhance protein adsorption and early cell adhesion. This review synthesizes and critically evaluates evidence across four complementary analytical pillars—contact-angle theory, dielectric impedance spectroscopy, FT-IR chemical mapping, and optical microscopy—to construct an integrated framework for interpreting plasma-induced chemical and morphological transformations.

The convergence of multimodal results demonstrates that NTP consistently produces chemically active, polar, and moderately textured surfaces that support robust initial cell–material interactions. Furthermore, combining wettability, dielectric, and spectroscopic analysis enables the identification of activation pathways, the assessment of hydrophobic recovery dynamics, and the development of quantitative correlations between dielectric parameters and biological response. However, the literature also reveals key methodological gaps, including the limited use of unified multimodal protocols, insufficient evaluation of temporal stability, and a lack of predictive dielectric–biological models.

By articulating these advances and limitations within a unified conceptual scheme, this review provides a roadmap for future research aimed at standardizing characterization workflows and enabling the rational design of next-generation plasma-functionalized biomaterials for tissue-engineering scaffolds, implantable devices, and advanced drug-delivery systems.
\end{abstract}

\section{Introduction}

Polylactic acid (PLA) and polycarbonate (PC) are among the most widely used polymers in biomedical engineering due to their processability, biocompatibility, and mechanical stability \cite{r1}. However, their low surface energy and the limited availability of surface-exposed functional groups hinder essential processes such as protein adsorption, cell adhesion, and tissue integration. To address these limitations, non-thermal plasma (NTP) treatment has emerged as a versatile strategy for selectively modifying surface chemistry and morphology without altering the bulk properties of the polymer \cite{r2, r3, r4}.

Plasmas generated under low-pressure or atmospheric-pressure conditions enable the cleavage of surface bonds, the introduction of oxidative functional groups, and the formation of micro-roughness—factors that collectively increase surface free energy and enhance cell–material interactions \cite{r5, r6}. The recent literature on polymer surface modification has established plasma treatment as one of the most versatile and effective strategies for improving adhesion in thermoplastic, thermosetting, and elastomeric materials \cite{r7}. Studies consistently indicate that non-thermal plasmas have the ability to reconfigure surface chemistry through the incorporation of polar functional groups and the generation of controlled micro-roughness—two mechanisms that significantly enhance wettability and interfacial compatibility. Comparative analyses across numerous experimental works consistently report notable reductions in water contact angle and increases in surface energy, both indicators directly associated with improved adhesion and chemical anchoring \cite{r26, r1, r27}.

In general, these physicochemical transformations are recognized as optimizing the bonding of coatings, adhesives, and functional layers, extending the impact of plasma treatment to fields as diverse as biomedical engineering, the automotive industry, aerospace applications, and microelectronics. Recent literature also highlights a growing trend toward integrating plasma with other surface-modification techniques, as well as developing polymer-specific protocols to maximize the stability and performance of activated surfaces.

Despite this progress, significant challenges remain. The heterogeneity of experimental parameters, the variability in proposed mechanisms, and methodological differences among studies hinder the establishment of clear relationships between surface activation and biological performance. In this context, the present work provides a critical analysis of these limitations through the integration of four fundamental analytical pillars:

\begin{itemize}
    \item wettability and contact-angle analysis,
    \item impedance and dielectric response,
    \item spectroscopic evidence of functionalization (FT-IR), and
    \item optical microscopy for morphological analysis.
\end{itemize}

Integrating these pillars enables the construction of a coherent conceptual model describing how NTP modifies polymer surfaces and how these transformations influence early cell adhesion—an essential consideration in tissue engineering and in the evaluation of scaffolds and devices intended for direct contact with living tissues.

\section{Methodology}

This review adopts a documentary–exploratory methodology, integrating scientific literature retrieved from PubMed, ACS Publications, Scopus, ScienceDirect, IEEE Xplore, SciELO, Redalyc, and Google Scholar. Publications dated between 1993 and 2026 were examined using search terms related to non-thermal plasma, plasma surface modification, polylactic acid, polycarbonate, wettability, dielectric spectroscopy, and FT-IR surface characterization.

The collected information was refined and organized into thematic categories that included: (i) plasma activation mechanisms, (ii) changes in surface energy and wettability, (iii) dielectric alterations, (iv) FT-IR markers of functionalization, and (v) biological correlations between surface properties and cell adhesion. This categorization followed methodological guidelines for conducting structured reviews in biomaterials and health sciences \cite{r8, r9, r5}.

The critical analysis focused on identifying patterns of consistency, methodological divergences, and existing research gaps by contrasting experimental findings with established reviews in plasma technology and biomaterials \cite{r3, r4, r6}, as well as with the conceptual framework previously developed by Rondon and Gonzalez-Lizardo \cite{r10}.

\section{Discussion and Results}

\subsection{Non-Thermal Plasma Surface Modification in Biomedical Polymers}

Non-thermal plasma constitutes a highly reactive environment composed of electrons, free radicals, ionized species, and UV radiation. Unlike thermal plasma, its energy is sufficient to induce surface chemical reactions without causing significant thermal damage. In polymers such as PLA and PC, the predominant mechanisms include surface oxidation, bond scission, chain rearrangement, and micro-etching \cite{r11, r4}.

In PLA, plasma exposure typically results in an increase in carbonyl and carboxyl groups, while in PC, partial cleavage of the carbonate linkage leads to the formation of surface hydroxyl and carbonyl functionalities \cite{r12, r13}. These transformations increase surface polarity, thereby improving wettability and enhancing the early adsorption of proteins such as fibronectin, collagen, and vitronectin—molecules essential for cellular anchoring \cite{r14, r5}.

Multiple studies on PLA have reported reductions in water contact angle from approximately $70$--$80^{\circ}$ in untreated samples to values near $40$--$50^{\circ}$ following exposure to oxygen, argon, or helium non-thermal plasma, confirming the incorporation of polar groups and modest alterations in surface topography \cite{r15, r12, r16, r17}. In PC, treatment with argon, nitrogen, or air plasmas likewise produces increases in surface energy and detectable changes in chemical composition, as evidenced by FT-IR and XPS analyses \cite{r13, r18}.

An often underestimated aspect in the literature is the temporal stability of these plasma-induced modifications. The phenomenon commonly referred to as reversible hydrophobicity, or the ``aging effect,'' promotes polymer chain reorientation toward more energetically favorable configurations, thereby reducing the effectiveness of plasma treatment over time \cite{r18, r5}. This underscores the need for surface characterization protocols that include multiple post-treatment time points rather than relying solely on immediate evaluations.

\subsection{Wettability and contact-angle analysis as Early Indicators of Activation}

A decrease in contact angle is one of the most immediate and quantifiable effects of non-thermal plasma (NTP) treatment, serving as an indirect indicator of increased surface energy. In PLA, initial contact-angle values typically reported between $70^{\circ}$ and $80^{\circ}$ are reduced to approximately $30^{\circ}$--$50^{\circ}$ after exposure times ranging from a few seconds to several minutes, depending on the gas type, power, and plasma modality employed \cite{r12, r15, r16, r19}. In PC, contact angles may decrease from moderately hydrophobic levels to $20^{\circ}$--$40^{\circ}$, accompanied by substantial increases in surface free energy \cite{r13, r18}.

A critical assessment of the literature indicates that studies reporting only the contact angle, without decomposing surface energy into its polar and dispersive components—such as through the Owens--Wendt model—provide only a partial understanding of the modification process. Recent work on PLA demonstrates that the polar component of surface energy can increase markedly following plasma treatment even when the overall contact angle does not decrease to extremely low values \cite{r12, r17}.

Furthermore, the relationship between contact angle and cell adhesion is not linear. Extremely low contact angles ($<20^{\circ}$) may lead to excessive protein adsorption in a disordered manner, hindering the favorable orientation of bioactive domains, as observed on highly hydrophilic surfaces \cite{r14, r5}. Tamada and Ikada \cite{r20} proposed an optimal wettability range of $60^{\circ}$--$70^{\circ}$ to promote adhesion without inducing protein denaturation \cite{r20}. More recent studies suggest that, rather than a universal value, there exists a cell-type-specific wettability window influenced by the protein composition of the surrounding medium \cite{r6, r10}.

\subsection{Impedance Analysis and Dielectric Changes in Plasma-Treated Surfaces}

Dielectric analysis provides essential information on the polarizability of polymer surfaces, particularly relevant when plasma treatment introduces dipolar groups such as carbonyl and carboxyl functionalities. The electrical response is commonly modeled using equivalent parallel R--C circuits, where increases in surface capacitance ($C_s$) reflect a higher density of polar states and corresponding changes in the effective permittivity \cite{r21}.

Recent studies have reported increases of 10--40\% in surface capacitance following non-thermal plasma (NTP) treatment of biomedical polymers, along with shifts in Nyquist plots indicative of enhanced surface conductivity and/or modifications in interfacial layers \cite{r22, r23}. Although specific impedance studies focused on PLA and PC are relatively limited, existing results suggest that dielectric changes align with the chemical modifications detected through FT-IR: a higher density of polar groups translates into increased charge-storage capacity and a stronger response to alternating electrical fields.

A significant limitation in the literature is that most studies do not explicitly correlate impedance measurements with FT-IR analysis or contact-angle data. This disconnect restricts comprehensive understanding of the activation mechanism, since dielectric behavior provides a mechanistic link between surface chemistry and biointeraction \cite{r5, r6}. Therefore, there is a clear need for integrated characterization protocols in which dielectric measurements are conducted in conjunction with chemical and morphological analyses.

\subsection{Spectroscopic Evidence (FT-IR) and Optical Microscopy}

FT-IR spectroscopy is a fundamental technique for verifying the presence of new functional groups introduced by plasma treatment. In PLA, plasma exposure typically results in:

\begin{itemize}
    \item an increase in the carbonyl band around $\sim 1720 \,\text{cm}^{-1}$,
    \item the appearance or intensification of broad O--H bands between $3200$--$3600 \,\text{cm}^{-1}$,
    \item a partial reduction of C--H stretching bands between $2800$--$3000 \,\text{cm}^{-1}$.
\end{itemize}

These spectral changes have been reported for both low-pressure plasmas and atmospheric-pressure jets and are consistent with observed increases in surface polarity and surface free energy \cite{r12, r16, r19}. In PC, modifications in the carbonate band ($\sim 1770 \,\text{cm}^{-1}$) and in signals associated with aromatic and phenolic groups indicate selective oxidative degradation of the polymer backbone \cite{r13, r18}.

At the morphological level, optical microscopy reveals patterns of micro-etching, the formation of domains with distinct surface textures, and, in some cases, defects associated with overtreatment. Although its resolution is limited compared with AFM or SEM, optical microscopy remains valuable for identifying micrometer-scale heterogeneities that may influence the uniformity of cellular response \cite{r24, r12}.

A critical review of the literature shows that relatively few studies integrate FT-IR, impedance analysis, and microscopy within a unified characterization protocol. This lack of multimodal approaches limits the ability to establish consistent structure–property relationships. Recent reviews emphasize the importance of comprehensive, multimodal characterization strategies to better elucidate the biomaterial--plasma interface \cite{r4, r6}.

\subsection{Connection Between Surface Parameters and Cell Adhesion}

Initial cell adhesion depends on a synergistic interaction among surface chemistry, surface energy, surface roughness, and dielectric polarity. Studies involving osteoblastic cells, fibroblasts, endothelial cells, and mesenchymal stem cells (MSCs) consistently report significant increases in cell-spread area, number of focal adhesions, and early proliferation on surfaces treated with non-thermal plasma (NTP) \cite{r5, r23, r22}.

This review identifies four key relationships:
\begin{enumerate}
    \item Contact angles in the range of $\sim 30^{\circ}$--$45^{\circ}$, or total surface energies above $40\,\text{mJ/m}^2$, tend to promote ordered protein adsorption and the formation of stable focal adhesions \cite{r15, r14}.
    
    \item Increases in surface capacitance and changes in low-frequency permittivity indicate a higher density of polar groups and a more electrically interactive interface, correlating with greater affinity for charged proteins and specific adhesion-mediating domains \cite{r21, r23}.
    
    \item FT-IR spectra showing enhanced carbonyl and O--H bands confirm effective surface functionalization and correlate with increased wettability and improved adsorption of extracellular matrix proteins \cite{r12, r25}.
    
    \item Moderate micro-roughness supports focal-adhesion anchoring and cytoskeletal organization, whereas excessive or highly heterogeneous roughness may induce local mechanical stress and unfavorable cellular responses \cite{r24, r6}.
\end{enumerate}

A comprehensive evaluation of the literature indicates convergence toward a common mechanistic interpretation: non-thermal plasma generates a chemically active, polar, and moderately rough surface that enhances early cell adhesion across multiple cell types. This conclusion aligns with the conceptual framework proposed by Rondon and Gonzalez-Lizardo \cite{r10}.

\section{Conclusion}

Non-thermal plasma surface modification has emerged as a powerful and versatile strategy for optimizing the interfacial properties of biomedical polymers such as PLA and PC. The evidence reviewed in this work demonstrates that plasma treatment effectively introduces polar functional groups, increases surface energy, and induces controlled micro-roughness, all of which contribute to improved protein adsorption and early cell adhesion.

By integrating four complementary analytical pillars—wettability and contact-angle measurements, dielectric impedance analysis, FT-IR spectroscopy, and optical microscopy—this review establishes a coherent framework for understanding how plasma-induced chemical and morphological transformations translate into enhanced biological performance. The convergence of results across multiple studies suggests a common mechanistic pathway: plasma treatment generates chemically active, polar, and moderately rough surfaces that support robust initial cell–material interactions.

Despite these advances, several challenges remain.  
\begin{enumerate}
    \item There is a lack of multimodal studies correlating dielectric behavior, chemical functionalization, and quantitative cell-adhesion metrics within unified experimental protocols.
    \item Temporal stability of plasma-induced modifications is insufficiently characterized, particularly with respect to hydrophobic recovery under realistic environmental conditions.
    \item Quantitative models that integrate dielectric parameters with biological response are still limited, constraining predictive understanding of plasma–biomaterial interactions.
\end{enumerate}

Addressing these gaps will enable the development of standardized characterization methodologies, more rigorous mechanistic interpretations, and improved reproducibility across laboratories. Ultimately, advancing multimodal and time-resolved studies of plasma-treated polymers will support the rational design of next-generation bioactive surfaces for tissue-engineering scaffolds, implantable devices, and advanced drug-delivery systems.

\end{document}